\documentclass[aps,pra,twocolumn,showpacs]{revtex4-1}
\usepackage{amsbsy,latexsym,amsmath}
\usepackage{amsfonts}
\usepackage{amssymb}
\usepackage[mathscr]{eucal}
\usepackage{epsfig,graphics,graphicx}
\usepackage{color}
\usepackage{float}

\newtheorem{lemma}{Lemma}

\begin{document}

 \title{A Geometric Approach to Quantum State Separation}
\author{ E. Bagan$^{1,2}$, V. Yerokhin$^{1}$, A. Shehu$^{1}$, E. Feldman$^{3}$, and J. A. Bergou$^{1}$}
\affiliation{$^{1}$Department of Physics and Astronomy, Hunter College of the City University of New York, 695 Park Avenue, New York, NY 10065, USA\\
$^{2}$F\'{i}sica Te\`{o}rica: Informaci\'{o} i Fen\`{o}mens Qu\`antics, Universitat Aut\`{o}noma de Barcelona, 08193 Bellaterra (Barcelona), Spain\\
$^{3}$Department of Mathematics, Graduate Center of the City University of New York, 365 Fifth Avenue, New York, New York 10016, USA }

\begin{abstract} 
Probabilistic quantum state transformations can be characterized by the degree of state separation they provide. This, in turn, sets limits on the success rate of these transformations. We consider optimum state separation of two known pure states in the general case where the known states have arbitrary {\emph{a priori}} probabilities. The problem is formulated from a geometric perspective and shown to be equivalent to the problem of finding tangent curves within two families of conics that represent the unitarity constraints and the objective functions to be optimized, respectively. We present the corresponding analytical solutions in various forms.  In the limit of perfect state separation, which is equivalent to unambiguous state discrimination, the solution exhibits a phenomenon analogous to a second order symmetry breaking phase transition. We also propose a linear optics implementation of separation which is based on the dual rail representation of qubits and single-photon multiport interferometry.
\end{abstract}
\maketitle

\section{Introduction}\label{sec intro}

Quantum information processing deals with changes in the state of a quantum system and what they amount to in terms of the information encoded in the initial state and the transformed or final state. Not any transformation from a given state, or set of states, to another is allowed by quantum mechanics, which sets strong limitations to the processing of information in quantum computation and quantum communication~\cite{Nielsen&Chuang}. Even so, quantum information processing is expected to outperform its classical counterpart~\cite{Bennett}. Some expectations are already materializing in quantum cryptography \cite{Crypto,Korzh} and quantum simulation \cite{simu,Thompson}, and much more is to come as experimentalists make progress overcoming decoherence and other issues involved in the implementation of quantum technologies.

Since the evolution of quantum states is central in quantum information, we are urged to investigate the ultimate limits imposed by nature on state transformations. In this sense, it has been recognized that probabilistic processing can offer significant advantages over deterministic processing. The simplest example is arguably unambiguous discrimination~\cite{Ivanovic,Dieks,Peres,Jaeger&Shimony}, which enables error-free identification of non-orthogonal quantum states, provided they are linearly independent \cite{CheflesLinInd,Bergourev}. More recently, perfect cloning has been proved possible when prior knowledge about the possible preparations of the state to be cloned is given \cite{DuanGuo,chclo,us1}. In all these cases, the price to pay is, of course, that the processing fails some times. However, we have means to know that the process has failed and we can compute the failure rate.

The examples above, as well as more recent developments in quantum replication \cite{Chiribella} and probabilistic metrology \cite{Gendra}, may be just the tip of the iceberg pointing at new directions in quantum processing.  In this paper we will focus on the simplest case of transformations over pure states belonging to a given two-state family. Such families are characterized by the overlap of their states and their prior probabilities. Any transformation acting on them gives a new two-state family also characterized by the overlap of the transformed states. Whether or not the transformation is possible with some given failure rate depends solely on the value of these overlaps and the prior probabilities of each of the states of the original family.  For zero failure rate, i.e., for deterministic processes the final overlap is necessarily larger or equal to the initial overlap. However, if some non-zero failure rate is allowed, the final overlap can be smaller than the initial overlap, and we say that the transformation increases the degree of separation~\cite{Chefles+Barnett} of the original states, since the final states become more easily distinguishable. Two states become fully separated under a transformation if the corresponding transformed states are orthogonal,  i.e., they have zero overlap. Full separation is equivalent to unambiguous discrimination in the following sense. If the transformed states are orthogonal, they can be  discriminated with no ambiguity by a projective measurement along the rays defined by the transformed states. So, the transformation can be used to implement unambiguous discrimination with the very same failure rate. Conversely, if two states can be unambiguously discriminated, upon identification we can prepare any state, in particular a state out of a pair with zero overlap. This shows that unambiguous discrimination followed by state preparation implement any transformation that fully separates two states. So, there is a measure-and-prepare protocol that implements any such transformation.

In intermediate situations, where some degree of separation is attained, there are several questions that we should answer. If the degree of separation is given, what is the optimal protocol, i.e., the protocol that has the smallest failure rate? If the failure rate cannot exceed a given value, what is the maximum degree of separation a transformation can attain as a function of the original overlap? And finally, what is the tradeoff between degree of separation and failure rate for a given initial overlap? These three questions are not independent, of course, but because of the impossibility  to find a fully explicit solution to separation, which would involve solving sixth degree polynomial equations, each one of them must be addressed separately. We provide the answers, i.e., the plots of the quantities relevant to each situation,  in a simple parametric form. This gives a full account of the separation problem.
The geometric approach, developed in \cite{us1} and \cite{Bergou1}, proves equally powerful here. It encompasses the entire physics in a simple intuitive picture and lends itself to analytical or numerical studies for which it provides a visual guidance.

A phenomenon analogous to a second order symmetry braking phase transition arises in the limit of full separation, i.e., when the overlap of the transformed states vanishes. This was already noticed in our recent letter~\cite{us1} on perfect cloning, which is a particular instance of separation since the overlap of the perfect clones is necessarily smaller than the overlap of the states to be cloned. There, we showed that the failure probability as a function of the prior probabilities is an analytic function if a finite number of clones are produced, but its second derivative becomes discontinuous in the limit of infinitely many clones. In this limit full separation takes place, since the overlap of the clones approaches zero exponentially as the number of clones increases. Here we show that such phase transition is a general feature of separation.

The paper is organized as follows. In Sec.~\ref{setup} we introduce the separation problem and our notation. We also show in detail that a unique solution exists. 
In Sec.~\ref{Q(eta)}, {we derive the minimum failure probability for a fixed degree of separation as a function of the prior probabilities. In the  particular case  of perfect cloning we recover the results of our previous work in~\cite{us1}. In Sec.~\ref{s' min} we derive the maximum separation for a fixed failure rate as a function of the initial overlap. In Sec.~\ref{tradeoff} we obtain the tradeoff curve between degree of separation and allowed failure probability. In Sec.~\ref{implementation} we provide a physical implementation based on single-photon, multiport interferometry employing the rail representation of qubits. We close with a brief discussion of our results in Sec.~\ref{conclusion}.

\section{Setup for quantum state separation}\label{setup}

We can always imagine that a probabilistic quantum transformation is carried out by a machine with an input port, an output port and two flags that herald the success or failure of the transformation.  The input $|\psi_i\rangle$, $i=1,2$ is fed through the input port for processing. In case of success, states~$|\psi'_i\rangle$, with the desired degree of separation, are delivered through the output port with conditioned probability~$p_i$. Otherwise, the output is in a failure state. Conditioned on the input state being $|\psi_i\rangle$, the failure probability is~$q_i=1-p_i$. 

We address optimality from a Bayesian viewpoint that assumes the states to be transformed are given with some {\it a priori}  probabilities $\eta_1$ and $\eta_2$, $\eta_1+\eta_2=1$. Then a natural cost function for our probabilistic machines is given by the average failure probability 
\begin{equation}
Q=\eta_1 q_1+\eta_2 q_2.
\label{obj fun}
\end{equation}
If $|\psi_i\rangle$ and the corresponding transformed states $|\psi'_i\rangle$ are given, the optimal machine is one that minimizes the cost function 
$Q$. In this case our aim is to find that optimal machine and the minimum average failure probability $Q_{\rm min}$ for arbitrary priors $\eta_1$ and $\eta_2$.

A different way of approaching optimality may consist in finding the machine (or machines) that achieves the highest degree of separation, namely, minimizes the overlap $s':=|\langle\psi'_1|\psi'_2\rangle|$ for given initial states $|\psi_i\rangle$, subject to the condition that the average probability~$Q$ does not exceed some given value, $Q_{\rm max}$. In this case we could further assume that either the initial overlap $s:=|\langle\psi_1|\psi_2\rangle|$ is given, in which case one can compute the tradeoff curve~$s'_{\rm min}(Q_{\rm max})$, or else assume that $Q_{\rm max}$ is fixed and compute the curve~$s'_{\min}(s)$. It is easy to see that $s'_{\rm min}(Q_{\rm max})$ and $Q_{\rm max}(s'_{\rm min})$ are just inverses of each other.

Whether we approach optimality one way or another depends merely on the problem at hand. Hence, e.g., for perfect cloning from one initial copy of either $|\psi_1\rangle$ or~$|\psi_2\rangle$ to $n$ final copies (i.e., $|\psi'_i\rangle=|\psi_i\rangle^{\otimes n}$), the former approach is most suitable since the final overlap is fixed, $s'=s^n$,  and so is the degree of separation attained by the cloner. So, in~\cite{us1} the solution was given in terms of~$Q_{\rm min}$ as a function of the prior probability $\eta_1$. However, one may need to know what is the maximum number of clones that can be produced if the failure rate cannot exceed~$Q_{\rm max}$, in which case one takes the latter approach, and computes~$n_{\rm max}=\log[s'(Q_{\rm max})]/\log s$. 

The machine that carries the probabilistic transformation is usually described by two Kraus operators $A_{\rm succ}$, $A_{\rm fail}$, so that~$A^\dagger{}_{\kern-.3em\rm succ}A_{\rm succ}+A^\dagger{}_{\kern-.2em\rm fail}A_{\rm fail}=\openone$~\cite{Nielsen&Chuang,Chefles+Barnett}. We can think of $A_{\rm succ}$ and $A_{\rm fail}$ as measurement operators. The transformation is successfully applied if the outcome of such (generalized) measurement is ``succ",  and fails otherwise.
Neumark's theorem provides an alternative approach  that turns out to be more convenient for our analysis.  Additional details on this method can be found in \cite{Bergou}. 
In this formulation, 
the Hilbert space ${\mathscr H}$ of the original states is supplemented with an ancillary space~${\mathscr H}_{\rm extra}\otimes {\mathscr H}_F$ that accommodates both the required extra-dimensions (if necessary) as well as the success/failure flags. Then, a unitary transformation~$U$ (time evolution) from ${\mathscr H}\otimes {\mathscr H}_{\rm extra}\otimes {\mathscr H}_{F}$ onto ${\mathscr H}'\otimes{\mathscr H}_F$ is defined through~\cite{DuanGuo,us1,chsep}
\begin{eqnarray}
U|\psi_1\rangle|0\rangle&=& \sqrt{p_1}|\psi'_1\rangle|\alpha_1\rangle +\sqrt q_1 |\phi\rangle|\alpha_0\rangle,\label{U1}\\
U|\psi_2\rangle|0\rangle&=& \sqrt{p_2}|\psi'_2\rangle|\alpha_2\rangle +\sqrt q_2 |\phi\rangle|\alpha_0\rangle. \label{U2}
\end{eqnarray}
Here the ancillas are initialized in a reference state~$|0\rangle$. The states of the flag associated with successful transformation~$| {\alpha_i}\rangle$ are constrained to be orthogonal to the state~$|\alpha_0\rangle$ that signals failure. Upon performing a projective measurement  on the flag space ${\mathscr H}_F$, the final state delivered through the output port of our probabilistic machine is either $|\psi'_i\rangle$, in case of success, or $|\phi\rangle$ in case of failure. So, the outcome of this measurement tells us if the machine has succeeded or failed in delivering the right transformed state.  On general grounds, optimality requires $|\alpha_1\rangle=|\alpha_2\rangle$. Here we choose to consider a more general setup where these two states are different
to include state discrimination, for which the success flag states must be fully distinguishable, so $\langle\alpha_1|\alpha_2\rangle=0$.
Likewise, we could consider an even more general setup with two failure states $|\phi_1\rangle$ and $|\phi_2\rangle$ in Eqs.~(\ref{U1}) and~(\ref{U2}). This is necessarily sub-optimal since we could probabilistically determine whether we received $|{\psi_1}\rangle$ or $|{\psi_2}\rangle$ by applying unambiguous discrimination to the failure states~$| {\phi_i}\rangle$.  Sometimes we would be  certain of the input state, in which case we could  prepare $|\psi'_1\rangle$ or $|\psi'_2\rangle$ accordingly,  thereby increasing the overall success rate.

Taking the inner product of Eqs.~(\ref{U1}) and ~(\ref{U2}) with themselves shows that our probabilities are normalized: $p_i+q_i=1$.
Similarly, by taking the product of Eq.~(\ref{U1}) with Eq.~(\ref{U2}), we find the unitarity constraint,
\begin{equation}
s=\sqrt{p_1 p_2}\, \beta+\sqrt{q_1 q_2},
\label{unit cond}
\end{equation}
where $\beta=s' |\langle \alpha_1|\alpha_2\rangle|$. Without any loss of generality, in deriving Eq.~(\ref{unit cond}) we have chosen
$\langle {\psi_1}|{\psi_2}\rangle$, $\langle {\psi'_1}|{\psi'_2}\rangle$ and~$\langle\alpha_1|\alpha_2\rangle$ to be real and positive.
We note that $0\le\beta\le s$, and~$\beta=0$ for both full separation ($s'=0$) and unambiguous discrimination ($ |\langle \alpha_1|\alpha_2\rangle|=0$), whereas for optimal separation~$|\langle \alpha_1|\alpha_2\rangle|=1$. If Eq.~(\ref{unit cond}) is satisfied, it is not hard to prove that~$U$ has a unitary extension on the whole Hilbert space and the Kraus operators,~\mbox{$A_{\rm succ}$, $A_{\rm fail}$,} can be obtained by tracing out the ancillary degrees of freedom.

Geometrically, Eq.~(\ref{obj fun}) defines a straight line in the $q_1$-$q_2$ plane for fix values of $Q$ and the priors. Using $p_i=1-q_i$, Eq. (\ref{unit cond}) defines curves in the same plane characterized by the values of $s$ and $\beta$. In Fig.~\ref{fig:1} we display these lines and  curves for representative values of the parameters. 
For convenient referencing, we gather in a lemma all the features of these curves that we will need.  Points (a)--(d) are straightforward, so only (e) and (f) are proven below.

\begin{figure}[H]
\vspace{2em}
\centering
\includegraphics[width=14em]{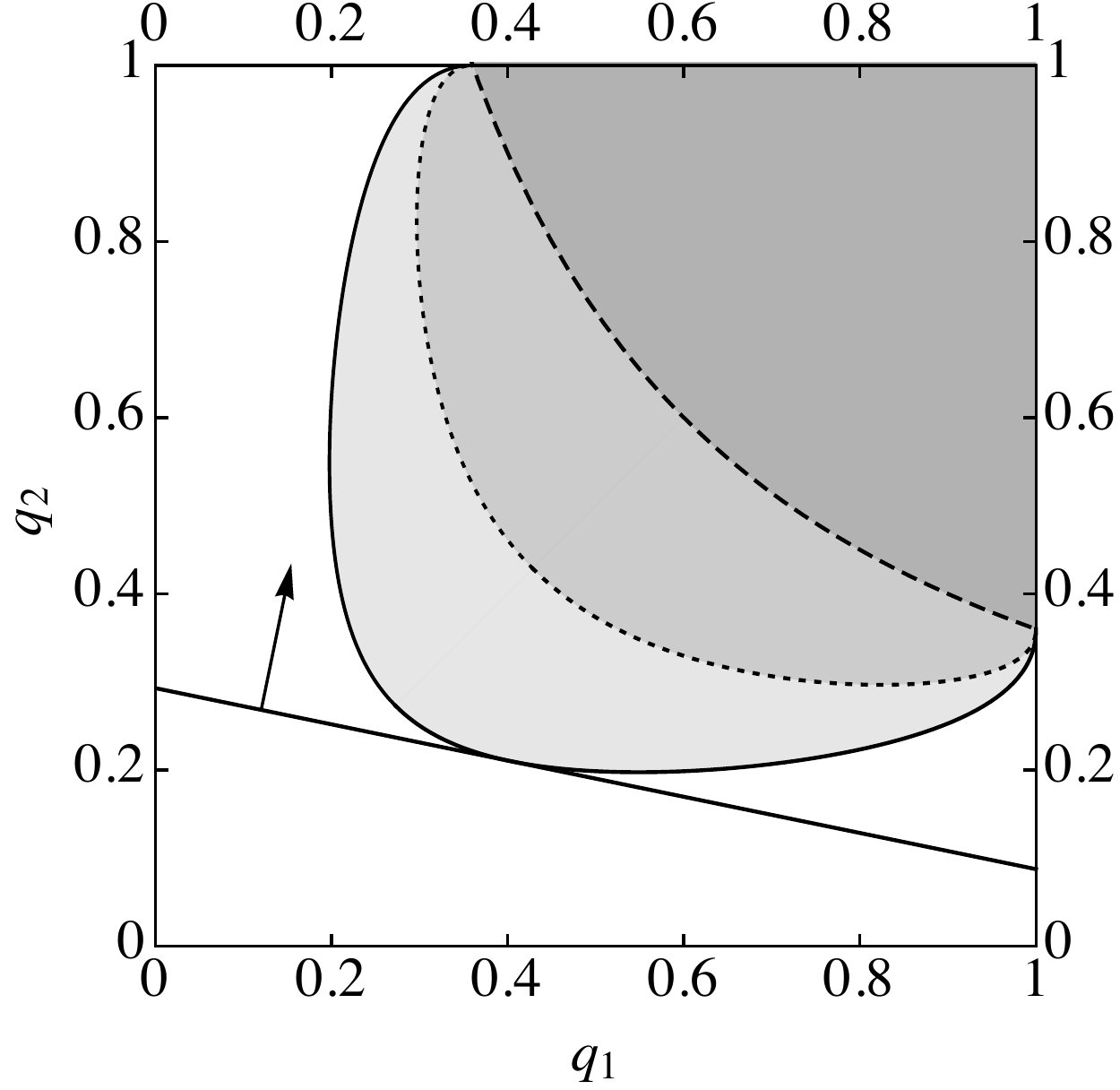}
\caption{Unitarity curves in Eq.~(\ref{unit cond}) and the associated sets~$S_\beta$ in Eq.~(\ref{S_alpha}) for $\beta=0.45$ (solid/light~gray), $\beta=0.30$ (dotted/medium gray), and $\beta=0$ (dashed/dark gray). The figure also shows the optimal straight segment \mbox{$Q=\eta_1 q_1+\eta_2 q_2$} and its normal vector~$(\eta_1,\eta_2)$. Plotted for  $s = 0.6$, $\eta_1=0.17$, $\eta_2=0.83$ and~$Q=0.24$.}
\label{fig:1}
\end{figure}

\begin{lemma}
(a)~For fixed $s$, Eq.~(\ref{unit cond}) defines a class of smooth curves on the unit square $0< q_i< 1$ (e.g., solid, dashed or dotted curves in Fig.~\ref{fig:1}). (b)~All these curves meet at their endpoints, $(1,s^{2})$ and $(s^{2},1)$. (c)~At the endpoints  the curves become tangent to the vertical and horizontal lines~$q_1=1$ and~$q_2=1$ respectively, provided~$\beta$ is not zero. (d)~For~$\beta=0$ the curve 
is an arc of the hyperbola $q_1 q_2=s^{2}$ (dashed line in Fig.~\ref{fig:1}). 
(e)~Each of these curves and the segments joining their end points with the vertex~$(1,1)$ enclose the sets (any of the gray regions in Fig.~\ref{fig:1})
\begin{equation}
S_\beta\!=\!\{ (q_1,\!q_2)\!\in\![0,\!1]\!\times\![0,\!1]\!: \sqrt{p_1 p_2}\,\beta+\sqrt{q_1 q_2}-s\!\ge\! 0\}.
\label{S_alpha}
\end{equation}
They satisfy $S_{\beta}\subset S_{\beta'}$ if $\beta<\beta'$.
(f)~Moreover, the sets~$S_\beta$ are convex. 
\end{lemma}

\noindent{\bf Proof.} (e)~The curve~(\ref{unit cond}) is readily seen to be part of the boundary of $S_\beta$. 
Assume that~$\beta\le\beta'$ and $(q_1,q_2)\in S_\beta$. Then 
\begin{equation}
\sqrt{p_1 p_2}\,\beta'+\sqrt{q_1 q_2}-s\ge \sqrt{p_1 p_2}\,\beta+\sqrt{q_1 q_2}-s\ge0,
\end{equation}
and thus $(q_1,q_2)\in S_{\beta'}$. (f)~To prove convexity let us assume that $(q_1,q_2)$ and $(q'_1,q'_2)$ belong to $S_\beta$. We define $\bar q_i=\lambda q_i+(1-\lambda)q'_i$, where $0\le\lambda\le1$. It follows that $\bar p_i:=1-\bar q_i=\lambda p_i+(1-\lambda)p'_i$. Since~$f(x,y)=\sqrt{xy}$ is a concave function in the unit square $\{(x,y)\, |\, 0\le x,y\le 1\}$, we have $\sqrt{\bar q_1\bar q_2}\ge \lambda \sqrt{q_1q_2}+(1-\lambda)\sqrt{q'_1q'_2}$ and, since~$\beta\ge0$, $\sqrt{\bar p_1\bar p_2}\,\beta\ge \lambda \sqrt{p_1p_2}\,\beta+(1-\lambda)\sqrt{p'_1p'_2}\,\beta$. Then
\begin{widetext}
\begin{equation}
\sqrt{\bar p_1 \bar p_2}\,\beta+\sqrt{\bar q_1 \bar q_2}-s\ge \lambda\left(\sqrt{p_1p_2}\beta+\sqrt{q_1q_2}-s\right)+(1-\lambda)\left(\sqrt{p'_1p'_2}\beta+\sqrt{q'_1q'_2}-s\right)\ge 0.
\end{equation}
\end{widetext}
Thus, $(\bar q_1,\bar q_2)\in S_\beta$, which proves the convexity of $S_\beta$ for~$\beta\ge0$.
$\blacksquare$

 Now that we have characterized the geometry of the unitarity constraint, a geometrical picture of the optimization problem emerges (See Fig.~\ref{fig:1}). 
Eq.~(\ref{obj fun}) defines a straight segment  on the square $0\le q_i\le 1$ with a normal vector in the first quadrant parallel to $(\eta_1,\eta_2)$. For fixed {\em a priori} probabilities, the average failure probability~$Q$ is proportional to the distance from this segment to the origin~$(0,0)$. The intersection of such a straight segment 
with the boundary 
of $S_\beta$ provides an admissible unitary transformation $U$ and its corresponding failure probability $Q$.
Since $S_{\beta}$ is convex and the stretch of its boundary given by Eq.~(\ref{unit cond}) is smooth, the optimal transformation, for which $Q$ is minimal, is defined by the unique point~$(q_1,q_2)$ of tangency with the segment~(\ref{obj fun}) that exists for any value of the priors and for $\beta>0$.
So, this tangency point determines the minimum failure probability $Q_{\rm min}$ and defines the optimal separation strategy through Eqs.~(\ref{U1}) and~(\ref{U2}). 

For $\beta=0$ (full separation/unambiguous discrimination),  the right hand side of Eq.~(\ref{unit cond}) describes a hyperbola for a fixed value of $s$, $q_2=s^2/q_1$, corresponding to a dashed line in Fig.~\ref{fig:1}. Its  slope, $q'_2=-s^2/q_1^2$, is in the range~$[-s^{-2},-s^{2}]$.  A unique point of tangency with the line~(\ref{obj fun}) can only exists if the slope of this line,~$-\eta_1/\eta_2$, is within this same range, namely if $s^2/(1+s^2)\le \eta_1\le1/(1+s^2)$.
The tangency point is then seen to be $(q_1,q_2)=\sqrt{\eta_1\eta_2}\,s\, (\eta^{-1}_1,\eta^{-1}_2 )$. This leads to a minimum average failure probability given by $Q_{\rm min}=Q_{\rm UD}:=2\sqrt{\eta_1\eta_2}s$, where the subscript UD stands for unambiguous discrimination.
If the slope is outside the range tangency is not possible, and then the optimal line merely touches the end points of the hyperbola. For $\eta_1<s^2/(1+s^2)$, the straight segment~(\ref{obj fun}) pivots on the lower end point, $(1,s^2)$, as we vary~$\eta_1$ and we have the minimum average failure probability as $Q_{\rm UD}=\eta_1+\eta_2 s^2$. Likewise, for $\eta_1>1/(1+s^2)$, the pivoting point is the upper end point of the hyperbola, $(s^2,1)$, which leads to~$Q_{\rm UD}=\eta_1 s^2+\eta_2$. The above can be summarized by
\begin{equation}
Q_{\rm UD}=\left\{
\begin{array}{ll}
2\sqrt{\eta_1\eta_2}\, s,&\displaystyle \frac{s^{2}}{1+s^{2}}\le\eta_1\le \frac{1}{1+s^{2}};\\[.8em]
\eta_1+s^{2} \eta_2, \quad &\displaystyle 0\le \eta_1\le \frac{s^{2}}{1+s^{2}};\\[.8em]
\eta_1 s^2+ \eta_2, \quad &\displaystyle \frac{1}{1+s^{2}}\le\eta_1\le  1.
\end{array}
\right.
\label{UD}
\end{equation}
This expression reproduces the optimal average failure probability for unambiguous 
discrimination \cite{Jaeger&Shimony}, as it should.

Furthermore, we note that for the second (third) line in~(\ref{UD}) we have $p_1=1-q_1=0$ ($p_2=1-q_2=0$), which leads to a $2$-outcome projective measurement, as only the success flag state $|\alpha_2\rangle$  ($|\alpha_1\rangle$) is needed in Eqs.~(\ref{U1}) and~~(\ref{U2}). The solution in the first line of Eq.~(\ref{UD}) is manifestly symmetric under the exchange of the input states, i.e., under~$\eta_1\leftrightarrow\eta_2$. However, this symmetry is lost in the other lines. Instead,  the effect of swapping the states turns the solution in the second line of Eq.~(\ref{UD}) into the solution in the third line.  One can also check that~$Q_{\rm UD}$ is a twice differentiable function of $\eta_1$ (or $\eta_2$), with a second derivative discontinuous at~$\eta_1=s^2/(1+s^2)$ and $\eta_1=1/(1+s^2)$. Our geometrical approach shows that the average failure probability $Q_{\rm min}$ is an infinitely differentiable function of~$\eta_1$ for $\beta>0$, since according to our lemma, the boundary curve~(\ref{unit cond}) merges smoothly into the lines $q_1=1$ and $q_2=1$. 
So, it turns out that at $\beta=0$ a phenomenon similar to a second order symmetry breaking phase transition takes place. A~similar phenomenon was observed in unambiguous discrimination of more than two pure states~\cite{Bergou1}. 

Our lemma can likewise be used to address optimality for given priors~$\eta_1$ and~$\eta_2$ and average failure probability not exceeding~$Q_{\rm max}$, with $0\le Q_{\rm max}< Q_{\rm UD}$. First, since the unitarity curve is a function of~$\beta=s' \langle\alpha_1|\alpha_2\rangle$, we set $|\alpha_1\rangle=|\alpha_2\rangle$, i.e., $\langle\alpha_1|\alpha_2\rangle=1$, to ensure the minimum value of $s'>0$ for a given $\beta$. Then, it follows from the lemma that the minimum final overlap $s'>0$ (the maximum degree of separation attainable), which we call~$s'_{\rm min}$, is that for which the segment~(\ref{obj fun}), with $Q=Q_{\rm max}$, and the boundary of $S_{\beta=s'}$ become tangent. Setting the margin $Q_{\rm max}$ in the  range~$[Q_{\rm UD}, 1]$ leads, obviously, to the trivial solution $s'_{\rm min}=0$, for such margin would allow full separation using unambiguous discrimination with a failure rate of exactly $Q=Q_{\rm UD}$, below the given margin.

In summary, our lemma provides the solution to optimal state separation from a geometrical viewpoint by showing that it is a convex optimization problem, for which a unique solution exists. Unfortunately,  a closed form for this solution does not exist for arbitrary prior probabilities, since finding the tangency point of the segment in Eq.~(\ref{obj fun}) with the curve in Eq.~(\ref{unit cond}) requires solving a six degree polynomial equation, as one can easily check. In the next sections, we give an analytic solution to state separation in parametric form. This solution contains all the information one may need in a simple and straightforward fashion. In particular, it enables us to easily draw plots of the relevant quantities for the various cases we will consider. 

\section{Minimum failure probability for a fixed degree of separation}\label{Q(eta)}

When the overlap of the final states is fixed, as in perfect cloning, we argued above that a natural problem consists in deriving the minimum failure rate of the optimal protocol, $Q_{\rm min}$, as a function of one of the priors, say $\eta_1$. In this section we address this problem by following the method employed in our derivation for cloning  in~\cite{us1}. All the expressions below can be obtained from their analogs in~\cite{us1} with the simple replacements~$s^m\to s$ and~$s^n\to s'$, starting with
the symmetric parametrization of the curve~(\ref{unit cond}). Its lower half (for which $q_2\le q_1$) is parametrized as
\begin{equation}
q_i={1-xy-(-1)^i\sqrt{1-x^2}\sqrt{1-y^2}\over2} ,\quad i=1,2,
\label{par sqrt}
\end{equation}
where
\begin{equation}
x={1-(1+s')t\over s'/s},\qquad y={1-(1-s')t \over s'/s}.
\label{x & y}
\end{equation}
This parametrization arises from a change of variables that linearizes the unitarity constraint, which proved very convenient in~\cite{us1}, where the advantages of its highly symmetric form were also apparent. The upper half of the curve~(\ref{unit cond}) can be obtained by applying the transformation $q_1\leftrightarrow q_2$. However, without any loss of generality, we can assume that $0\le \eta_1\le 1/2$  (thus, $1/2\le\eta_2\le1$), so only the lower half given by Eq.~(\ref{par sqrt}) can actually become tangent to the straight segment in Eq.~(\ref{obj fun}). 

Fig.~\ref{fig:2}~(a) shows plots of the unitarity curve [Eq.~(\ref{par sqrt}) plus the reflection $q_1\leftrightarrow q_2$] for $s=0.6$ and $s'=0.05$, $0.3$, $0.5$ and~$0.59$. For $s'=0.59$, very close to the value of $s$ (small separation), the vertex of the curve approaches the origin, which becomes a singular point in the limit \mbox{$s'\to s$}. As~$s'$ decreases (increasing separation), the curves approach the hyperbola $q_1 q_2=s^2$. It is apparent from the figure that the curves merge smoothly onto the lines~$q_1=1$ and~$q_2=1$ for the larger values of~$s'$. It becomes less obvious for small values of~$s'$, such as $s'=0.05$. However a blowup of Fig.~\ref{fig:2}~(a) would reveal that this is so. A~cusp at $(s^2,1)$ and $(1,s^2)$ arises only for $s'\to0$.

It follows from our lemma, and it can be checked using  Eq.~(\ref{par sqrt}), that the slope of the lower half of the unitarity curve increases monotonically as we move away from the line~$q_1=q_2$, where it has the value~$-1$, and vanishes before we reach the line~$q_1=1$. The values of $t$ at which the slope is $-1$ and $0$ are, respectively,
\begin{equation}
t_{-1}={1-s'/s\over 1-s'},\quad
t_0={1-s'{}^2/s^2\over 1-s'{}^{2}}.
\label{t's}
\end{equation}
So, there is a straight segment~(\ref{obj fun}), with slope $-\eta_1/\eta_2$, that is tangent to each point $(q_1(t),q_2(t))$, $t\in[t_{-1},t_0]$, of the unitarity curve parametrized by Eq.~(\ref{par sqrt}). Since the slope of this curve is $q'_2(t)/q'_1(t)$, where the prime stands for derivative with respect to $t$, a parametric expression for $\eta_1$ can be obtained from the equal slope condition~$-\eta_1/\eta_2=q'_2(t)/q'_1(t)$. The parametric expression for $Q_{\rm min}$ follows from imposing that $(q_1(t),q_2(t))$ must be a point of the straight segment~(\ref{obj fun}), so $Q_{\rm min}=\eta_1q_1(t)+\eta_2 q_2(t)$.
The final result can be cast as
\begin{equation}
\eta_1={q'_2\over q'_2-q'_1},\;\; Q_{\rm min}={q'_2 q_1-q'_1 q_2\over q'_2-q'_1},\;\; t_{-1}\le t\le t_0,
\label{main}
\end{equation}
where we have dropped the argument of $q_i(t)$ and $q'_i(t)$ to simplify the equation. Further, one can check that the derivatives of $q_i(t)$ can be written as
\begin{equation}
q'_i={\sqrt{q_i(1-q_i)}\over s'/s}\left\{{1+s'\over\sqrt{1-x^2}}-(-1)^i{1-s'\over\sqrt{1-y^2}}\right\}.
\end{equation}
Eq.~(\ref{main}) gives $Q_{\rm min}(\eta_1)$ in parametric form for $0<s'<s$. The solution for $s'=0$ was already derived in the previous section and for $s'=s$ we have the trivial solution $Q_{\rm min}=0$. These special cases can also be derived from Eq.~(\ref{main}) by carefully taking the corresponding limits.
The values of~$Q_{\rm min}$ at the end points of this range follow by substituting $t_{0}$ and $t_{-1}$, Eq.~(\ref{t's}), into Eq.~(\ref{par sqrt}). They are given by
\begin{equation}
Q_{0}=q_2(t_0)={s^{2}-s'{}^{2}\over 1-s'{}^{2}},\quad
Q_{-1}={s-s'\over 1-s'},
\label{Q's}
\end{equation}
where $Q_{\rm min}=Q_{-1}$ holds for equal priors and $Q_{\rm min}=Q_0$ for $\eta_1\to 0$ (i.e., $\eta_2\to 1$).

Fig.~\ref{fig:2}~(b) shows plots of the curves $Q_{\rm min}(\eta_1)$ for the same values of $s$ and $s'$ as the ones given above (solid lines). We see that $Q_{\rm min}$ is an increasing function of $\eta_1$ in the given range $[0,1/2]$, as one should expect. The figure also shows the failure rate for unambiguous discrimination (dashed line), which coincides with $Q_{\rm min}$ for~$s'=0$. From the plots, it is clear that $Q_{\rm min}$ is a decreasing function of~$s'$, again as it should be. 
\begin{figure}[t]
\centering
$%
\begin{array}{c}
\includegraphics[width=26em]{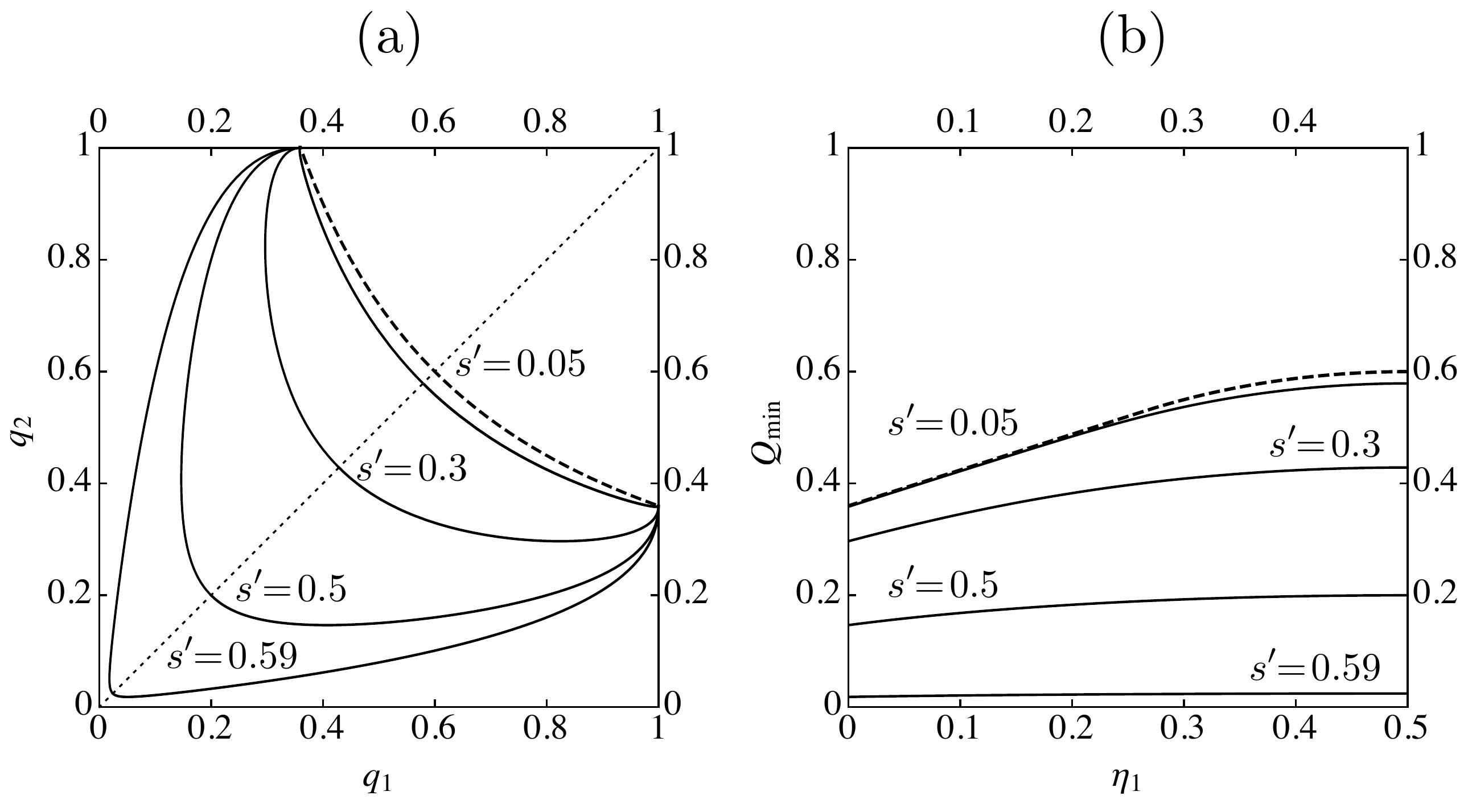}\\
\end{array}%
$%
\caption{(a) Unitarity curves for different values of~$s'$. The curves are symmetric under mirror reflexion along the  (dotted) straight line $q_1=q_2$, i.e., under the transformation~\mbox{$q_1\leftrightarrow q_2$}.  (b) Minimum separation failure probability~$Q_{\rm min}$ vs. $\eta_1$ (solid lines), for the same values of $s'$ used in~(a). In both figures the dashed lines correspond to full separation/unambiguous discrimination ($s'=0$) and  the value of the initial overlap is $s=0.6$.}
\label{fig:2}
\end{figure}

\section{Maximum separation}\label{s' min}

In this section, we assume that $\eta_1$, $\eta_2$ are fixed given quantities and we focus on the relationships among the initial overlap, the final overlap and the maximum allowed failure rate. To find the explicit form of these relationships, we will need to develop a new geometric view of both the unitarity constraint, Eq.~(\ref{unit cond}), and $Q= \eta_1 q_1+\eta_2 q_2$. 
We aim at a geometric representation simple enough to grasp visually the solution and yet powerful enough to provide this solution analytically. We show below that the unitarity curve and the straight segment of the previous sections can be mapped into conic curves, in particular into families of parabolas and ellipses respectively. This is arguably the simplest extension to our geometric description of state separation. The desired transformation, similar in spirit to that in~\cite{Roa}, is defined in terms of the new variables $u$ and $v$ as
\begin{equation}
u=\sqrt{q_1 q_2};\quad v={q_1+q_2\over 2}.
\label{transf}
\end{equation}
They are just the geometric  and arithmetic means of the failure probabilities, $q_1$ and $q_2$. Under this transformation the unitarity constraint becomes a parabola that can be conveniently written as
\begin{equation}
v={1+u^2\over2}-{(u-s)^2\over2s'^2}.
\label{unit cond conic}
\end{equation}
From this expression, one can immediately check that as~$s$ varies we obtain a family of parabolas whose envelope is yet another parabola, $v=(1+u^2)/2$, independently of $s'$. As $s'$ decreases from its maximum value $s'=s$, the parabolas in Eq.~(\ref{unit cond conic}) become thinner. For $s'=0$ they degenerate into the vertical segment $u=s$, $0\le v\le (1+s^2)/2$. These features are illustrated in~Fig.~\ref{fig:3}~(a).

Under the same transformation, Eq.~(\ref{transf}), the line $Q=\eta_1 q_1+\eta_2 q_2$ becomes an ellipse, which is most easily expressed parametrically in terms of the polar angle $\theta$, measured relative to the axis $v=0$ from the center of the ellipse. It is given by
\begin{eqnarray}
u&=&{Q\over\sqrt{1-\Delta^2}}\cos\theta,\nonumber\\
v&=&{Q\over1-\Delta^2}+{Q\Delta\over1-\Delta^2}\sin\theta,
\label{obj fun conic}
\end{eqnarray}
where we have defined $\Delta=\eta_2-\eta_1$.  
It is clear from this expression that the eccentricity of the ellipse is only a function of the priors. For equal priors, $\Delta=0$, the ellipse degenerates into the horizontal segment $v=Q$, $0\le u\le Q$, whereas for $Q=0$ it collapses into the origin $(u,v)=(0,0)$. As one increases $Q$, a family of similar ellipses is obtained. As they increase in size, their center moves up along the $v$ axis. The line $u=v$ is the envelope of this family, as one can easily check using~Eq.~(\ref{obj fun conic}). Fig.~\ref{fig:3}~(a) also illustrates these features.
\begin{figure}[t]
\centering
$%
\begin{array}{c}
\includegraphics[width=26em]{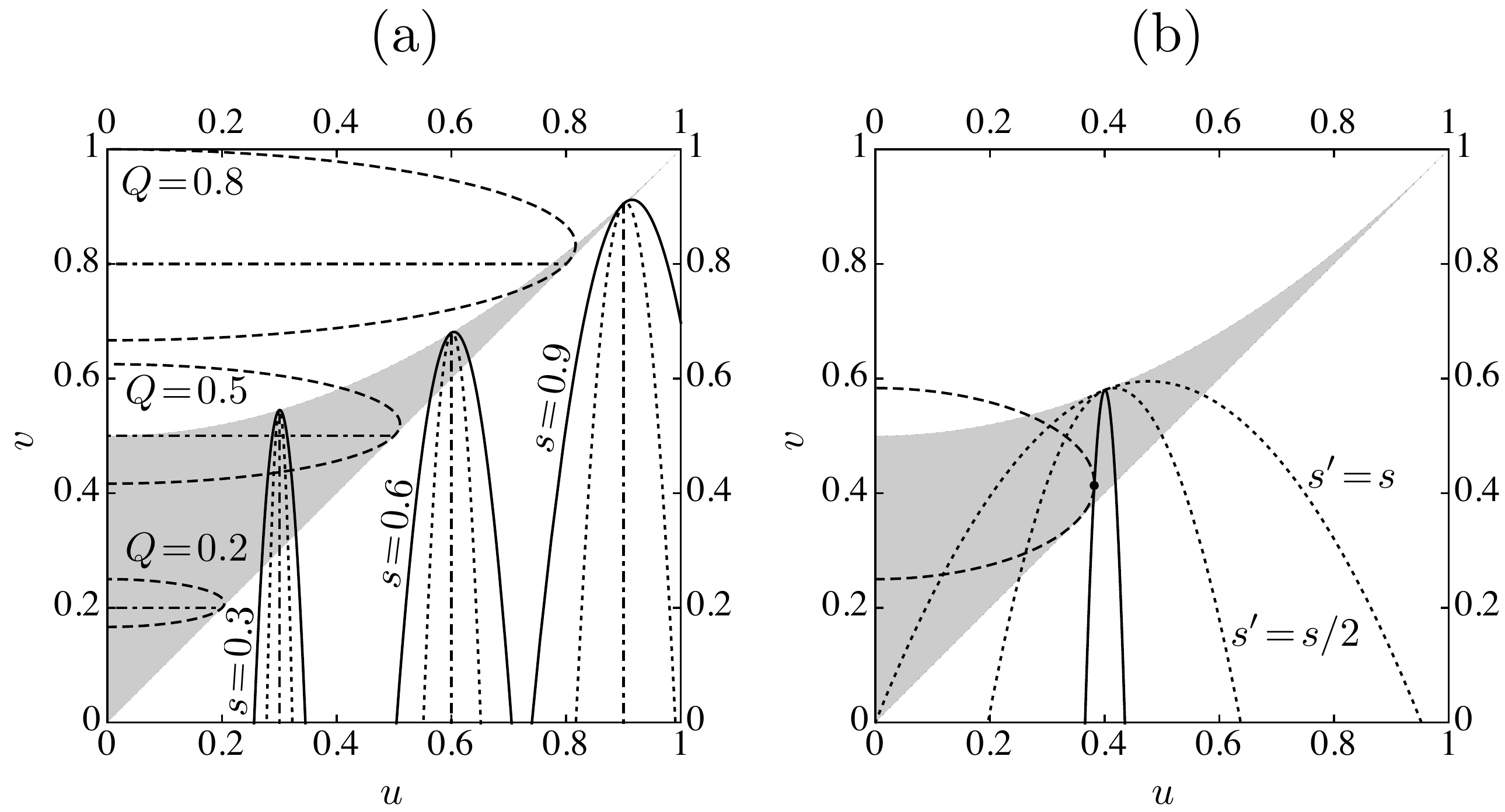}\\
\end{array}%
$%
\caption{(a) Unitarity parabolas, Eq.~(\ref{unit cond conic}), for different values of~$s$, $s'=s/7$ (solid lines) and $s'=s/14$ (dotted lines). The dashed lines are the ellipses in Eq.~(\ref{obj fun conic}) for various values of the failure rate $Q$. The top boundary line to the gray region, given by $v=(1+u^2)/2$, is the envelope of the solid and dotted parabolas. The bottom boundary line, i.e., the straight line~$v=u$, is the envelope of the family of ellipses (dashed lines). The geometric solution to optimal separation falls in the gray region. In this figure $\eta_1=0.4$. The degenerate curves for $s'=0$ (dot-dashed vertical line) and $\Delta=0$ (dot-dashed horizontal line) are also shown.   (b) Optimal (solid) and suboptimal (dotted) parabolas. The tangency point is also displayed. In this figure $\eta_1=0.3$, $s=0.4$ and $Q=Q_{\rm max}=0.35$. The optimal (minimum) value of $s'$, which gives the solid parabola, turns out to be $s'=0.032$. }
\label{fig:3}
\end{figure}

In terms of this conic geometry,  optimality is again given by a tangency point, this time between ellipses and parabolas. Because of the features of these families of conics, these points of tangency necessarily lie in the region between their envelopes,  which is the gray area in Fig.~\ref{fig:3}. Fig.~\ref{fig:3}~(b) illustrates optimality. Given a maximum failure rate $Q_{\rm max}$ and some initial overlap~$s$ ($Q_{\rm max}=0.35$ and $s=0.4$ in the example considered in the figure), we plot the corresponding ellipse defined by Eq.~(\ref{obj fun conic}) (dashed line). Among the various parabolas, characterized by the final overlap $s'$ (the figure shows two of them, for $s'=s$ and $s'=s/2$), the one that minimizes~$s'$ (solid line) has a unique point of tangency with the ellipse, thus giving us the solution, $s'_{\rm min}$.  To keep the notation simple we will drop the subscript ``$\rm min$'' wherever no confusion arises. 

To find the condition that gives the tangency point, we first note that the slopes of the ellipse and the parabolas are given respectively by
\begin{eqnarray}
{dv\over du}&=&{v'\over u'}=-{\Delta\over\sqrt{1-\Delta^2}}\cot\theta,\nonumber\\
{dv\over du}&=&u-{u-s\over s'^2},
\label{slopes}
\end{eqnarray}
where in the first line the primes stand for derivative with respect to the polar angle $\theta$. The right hand side of these two equations must be equal at the tangency point. Moreover, the tangency point must belong to both the ellipse and the optimal parabola. Hence
\begin{eqnarray}
Q{1\!+\!\Delta \sin\theta\over 1-\Delta^2}\!&=&\!{1\over2}\!+\!{Q^2\!\cos^2\theta\over2(1\!-\!\Delta^2)}
\!-\!{1\over2 s'^2}\!\!\left(\!{Q\cos\theta\over\sqrt{1\!-\!\Delta^2}}\!-\!s\!\!\right)^{\!\!2}\!\!,\nonumber
\\[.5em]
{\Delta\cot\theta\over	\sqrt{1-\Delta^2}}&=&{1-s'^2\over s'^2}{Q\cos\theta\over\sqrt{1-\Delta^2}}-{s\over s'^2}.
\label{eb21.04.15-2}
\end{eqnarray}
where to obtain the first (second) equation we have simply substituted Eq.~(\ref{obj fun conic}) into Eq.~(\ref{unit cond conic}) [Eq.~(\ref{slopes})].
Ideally, we would like to solve this system of equations by eliminating $\theta$, which would lead to a closed expression relating~$s$, $s'$ and $Q$. Unfortunately,  this involves solving a high degree polynomial equation in $\cos\theta$. Instead, 
we look at it as a system of two equations with two unknowns, $s$ and $s'$ (or $Q$ and $s'$) and keep $\theta$ as a parameter describing the curve $s'(s)$ [or $s'(Q)$] in parametric form. After some algebra, we obtain the simple expressions:
\begin{eqnarray}
s'\!&=&\!- {\sqrt{(1\!-\!Q)^2\!-\!(\Delta\!+\!Q\sin\theta)^2}\over \Delta\!+\!Q\sin\theta}\tan\theta,
\label{eb04.05.15-1}\\[1em]
s\!&=&\!{Q\Delta(1\!+\!\sin^2\theta)\!-\!(1\!-\!\Delta^2\!-\!2Q)\sin\theta\over\sqrt{1\!-\!\Delta^2}(\Delta\!+\!Q\sin\theta)\cos\theta} .
\label{eb04.05.15-2}
\end{eqnarray}
The range of values of the parameter $\theta$ in this equation is~$
-\arcsin\Delta\le\theta\le\theta_{\rm max}
$,
where
\begin{equation}
\theta_{\rm max}\!=\!\left\{
\!
\begin{array}{lll}
0&\mbox{if}&\displaystyle Q\le 1-\Delta,\\[.7em]
\displaystyle
\arcsin{1-Q-\Delta\over Q}
&\mbox{if}&\displaystyle Q\ge 1-\Delta.
\end{array}
\right.
\label{eb21.04.15-3}
\end{equation}
One can easily check the given minimum value of $\theta$ by substituting in Eqs.~(\ref{eb04.05.15-1}) and~(\ref{eb04.05.15-2}) to obtain $s'=s=1$, as it should be. Likewise, one can check that for $\theta=\theta_{\rm max}$ one has $s'=0$.  The two cases in Eq.~(\ref{eb21.04.15-3}) reveal the appearance of the phase transition in the limit $s'\to0$ that we discussed in previous sections. If $Q\ge1-\Delta$, substituting the second line of Eq.~(\ref{eb21.04.15-3}) in Eq.~(\ref{eb21.04.15-2}) we obtain $s=[(2Q+\Delta-1)/(1+\Delta)]^{1/2}$. Solving for $Q$, we find that $Q=\eta_1+s^2\eta_2$. This means that the condition $Q\ge1-\Delta$  is equivalent to $\eta_1+s^2\eta_2\ge 1-\Delta$, which can be immediately seen to give $\eta_1\le s^2/(1+s^2)$. So we obtain the second line in Eq.~(\ref{UD}), corresponding to the ``symmetry-broken phase". If $Q\le 1-\Delta$, namely, if $ s^2/(1+s^2)\le\eta_1$, we have instead $s=Q/\sqrt{1-\Delta^2}$. This equation can be written as $Q=2\sqrt{\eta_1\eta_2}s$. So, Eq.~(\ref{eb21.04.15-3}) has the same content as Eq.~(\ref{UD}). Recall that we are assuming $\eta_1\le1/2\le 1/(1+s^2)$. The third line in Eq.~(\ref{UD}) never applies under this assumption.

\begin{figure}[b] 
   \centering
\includegraphics[width=26em]{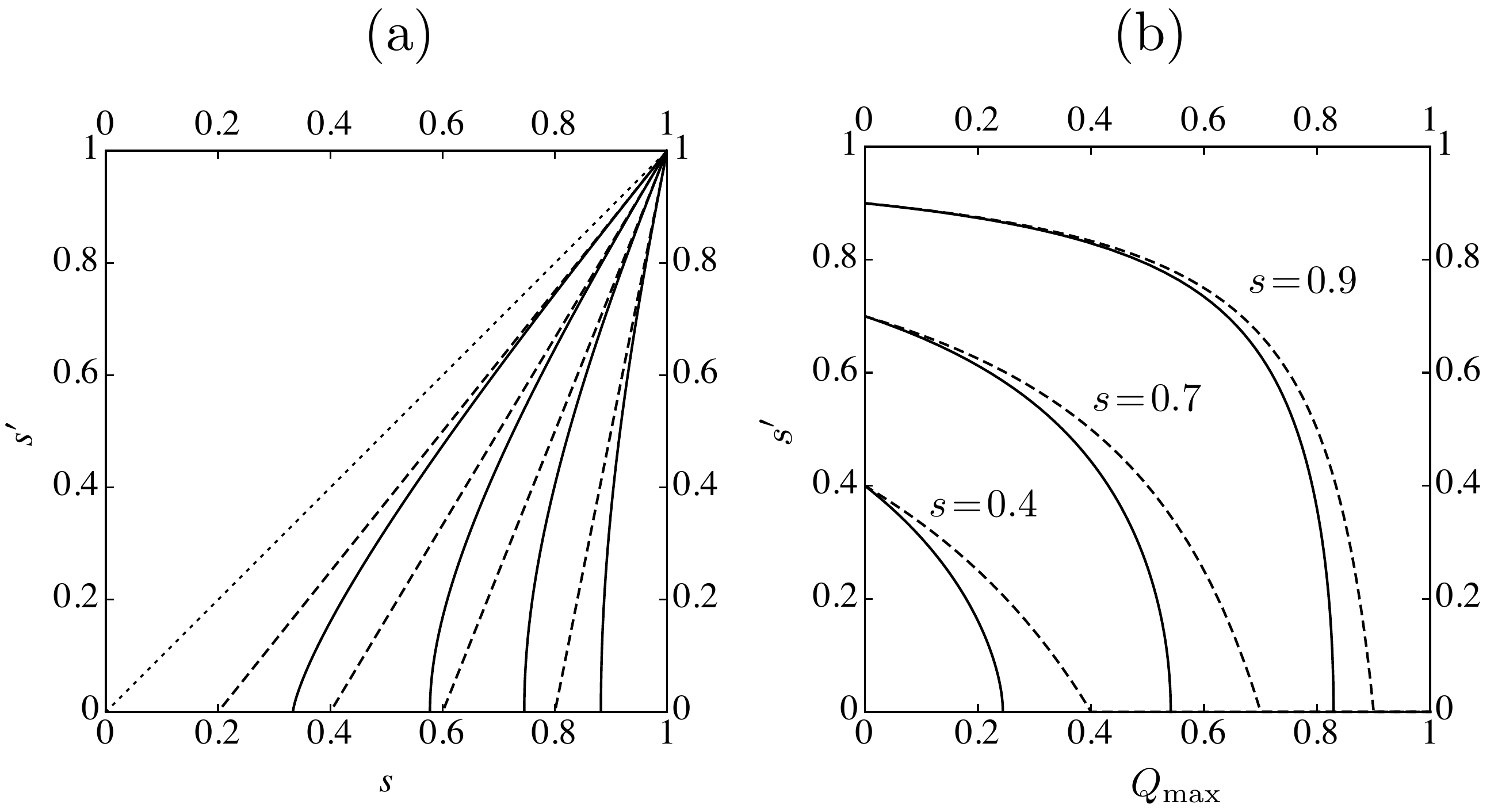}
   \caption{(a) Plots of $s'$ vs. $s$ for $\eta_1=0.1$ (solid lines) and $\eta_1=0.5$ (straight dashed lines) and for values of the failure rate. From left to right $Q_{\rm max}=0.2$, $0.4$, $0.6$, $0.8$. The dotted line is the (trivial) curve for $Q_{\rm max}=0$, which is the straight line $s'=s$. (b)~Minimum final overlap vs. maximum failure probability for various values of the initial overlap and the same two values of $\eta_1$ used in~(a).\vspace{-1em}
   }
   \label{fig:4}
\end{figure}

Eqs.~(\ref{eb04.05.15-1}) and~(\ref{eb04.05.15-2}) are plotted in Fig.~\ref{fig:4}~(a) for two possible priors: $\eta_1=0.1$ (solid lines) and $\eta_1=0.5$, i.e., for equal priors  (dashed lines). From left to right, the maximum allowed failure rate~$Q_{\rm max}$ is $0.2$, $0.4$, $0.6$ and~$0.8$. We see that for small values of the initial overlap, $s$, one can attain full separation ($s'=0$). Past the critical value,
\begin{equation}
s_{\rm cr}=
\left\{
\begin{array}{lll}
\displaystyle{Q_{\rm max}\over 2\sqrt{\eta_1\eta_2}}&\mbox{if}&\displaystyle Q\le 2\eta_1,\\[.9em]
\displaystyle
\sqrt{Q_{\rm max}-\eta_1\over\eta_2}
&\mbox{if}&\displaystyle Q\ge 2\eta_1,
\end{array}
\right.
\end{equation}
 full separation is no longer possible and~$s'$ increases (quite abruptly for small $\eta_1$). In the region $s<s_{\rm cr}$, the margin~$Q_{\rm max}$ is not saturated,  since the failure probability for unambiguous discrimination, $Q_{\rm UD}$, is  smaller than $Q_{\rm max}$. For $s\ge s_{\rm cr}$ we necessarily have to saturate the margin, i.e., $Q=Q_{\rm max}$. For equal priors (dashed lines) one can obtain the curves in explicit form from Eq.~(\ref{unit cond}) using that $q_1=q_2=Q$:
\begin{equation}
s'=
\left\{
\begin{array}{lll}
\displaystyle0 &\mbox{if}& s\le Q_{\rm max},\\[.2em]
\displaystyle
{s-Q_{\rm max}\over 1-Q_{\rm max}}
&\mbox{if}&\displaystyle s\ge Q_{\rm max}.
\end{array}
\right.
\label{tradeoff equal}
\end{equation}
This expression could also be obtained by carefully taking the limit $\Delta\to 0$ in Eqs.~(\ref{eb04.05.15-1})  through~(\ref{eb21.04.15-3}).
The figure clearly shows that separation becomes less demanding as we move away from the equal prior case. For \mbox{$Q_{\rm max}=0$}, i.e., in the deterministic limit, we recover the trivial solution $s'=s$ (dotted line).

\section{Tradeoff between Maximum separation and failure rate}\label{tradeoff}

By solving the system Eq.~(\ref{eb21.04.15-2}) for $Q$ and $s'$ , we obtain a parametric expression for the tradeoff curve $s'(Q)$ in terms of the polar angle $\theta$:
\begin{eqnarray}
s'^2\!&=&\!\sqrt{1\!-\!\Delta^2}\left({\sin\theta\over\Delta\!+\!\sin\theta}\right)^2\nonumber
\\
\!&\times&\!{\sqrt{1\!-\!\Delta^2}(1\!+\!s^2)\!\cos\theta\!-\!2s\!\left(1\!+\!\Delta \sin\theta\right)\over\cos\theta} ,
\label{tradeoff1}\\[.2em]
Q&=&\!{s\sqrt{1\!-\!\Delta^2}\!+\!\Delta\,s'^2\!\cot\theta\over(1\!-\!s'^2)\cos\theta}.
\label{tradeoff2}
\end{eqnarray}
Note that Eq.~(\ref{tradeoff1}) is an expression for the square of the final overlap. To keep the formula for~$Q$, Eq.~(\ref{tradeoff2}), 
short, we use~$s'^2$ as a shorthand for Eq.~(\ref{tradeoff1}). 
The range of $\theta$ in Eqs.~(\ref{tradeoff1}) and~(\ref{tradeoff2}) is:
$$
-\arctan\,{s\Delta\over\sqrt{1-\Delta^2}}\le\theta\le\theta_{\rm max},
$$
where the upper limit  of the interval can be written as
\begin{equation}
\theta_{\rm max}\!=\!\left\{
\!
\begin{array}{lll}
0&\mbox{if}&\displaystyle \eta_1\!\ge\!{s^2\over1\!+\!s^2},\\[.7em]
\displaystyle \!-\!\arccos\!{2s\sqrt{1-\Delta^2}\over1\!-\!\Delta\!+\!s^2(1\!+\!\Delta)}
&\mbox{if}&\displaystyle \eta_1\!\le\!{s^2\over1\!+\!s^2}.
\end{array}
\right.
\label{tradeoff cases}
\end{equation}
The lower limit  in the range of allowed $\theta$  can be derived from Eqs.~(\ref{tradeoff1}) and~(\ref{tradeoff2}) by imposing that $Q=0$ at $s'=s$. The upper limit can be derived from Eq.~(\ref{tradeoff1}) by imposing $s'=0$. Once again, we see that a second order phase transition occurs in the limit of full separation: by substituting the first (second) line of Eq.~(\ref{tradeoff cases}) in Eq.~(\ref{tradeoff2}) we obtain $Q=s\sqrt{1-\Delta^2}$\ ($Q=[1-\Delta+s^2(1+\Delta)]/2$), which is the first (second) case in Eq.~(\ref{UD}).

Fig.~\ref{fig:4}~(b) shows various plots of the separation vs.~$Q_{\rm max}$. As in Fig.~\ref{fig:3}, the plots are for~$\eta_1=0.1$ (solid lines) and for equal priors, $\eta_1=\eta_2=0.5$ (dashed lines). For equal priors, there is the explicit formula for the curves given in Eq.~(\ref{tradeoff equal}). Again, we see that as $\eta_1$ gets smaller, departing from the equal prior value~$1/2$, the states can be separated more for the same maximum rate of failure. As $Q_{\rm max}$ increases, the minimum overlap gets smaller, as it should. When~the margin~$Q_{\rm max}$ reaches the unambiguous discrimination value~$Q_{\rm UD}$ we have $s'=0$, attaining full separation. Larger values of~$Q_{\rm max}$ are rather meaningless in this context, since they will never be saturated by an optimal protocol, which requires  a failure rate of only $Q=Q_{\rm UD}$~($<Q_{\rm max}$) to fully separate the input states.

\section{A physical implementation: single-photon multiport interferometry}\label{implementation}

In this section we propose a physical implementation of optimal state separation. The implementation is based on the dual-rail representation of qubits and single-photon multiport interferometry using only linear optics elements, namely, a mirror and two beam splitters, BS1 and~BS2. The measurements are carried out by three photodetectors. The setup is sketched in Fig.~\ref{fig:5}.

\begin{figure}[t] 
   \centering
\includegraphics[width=18em]{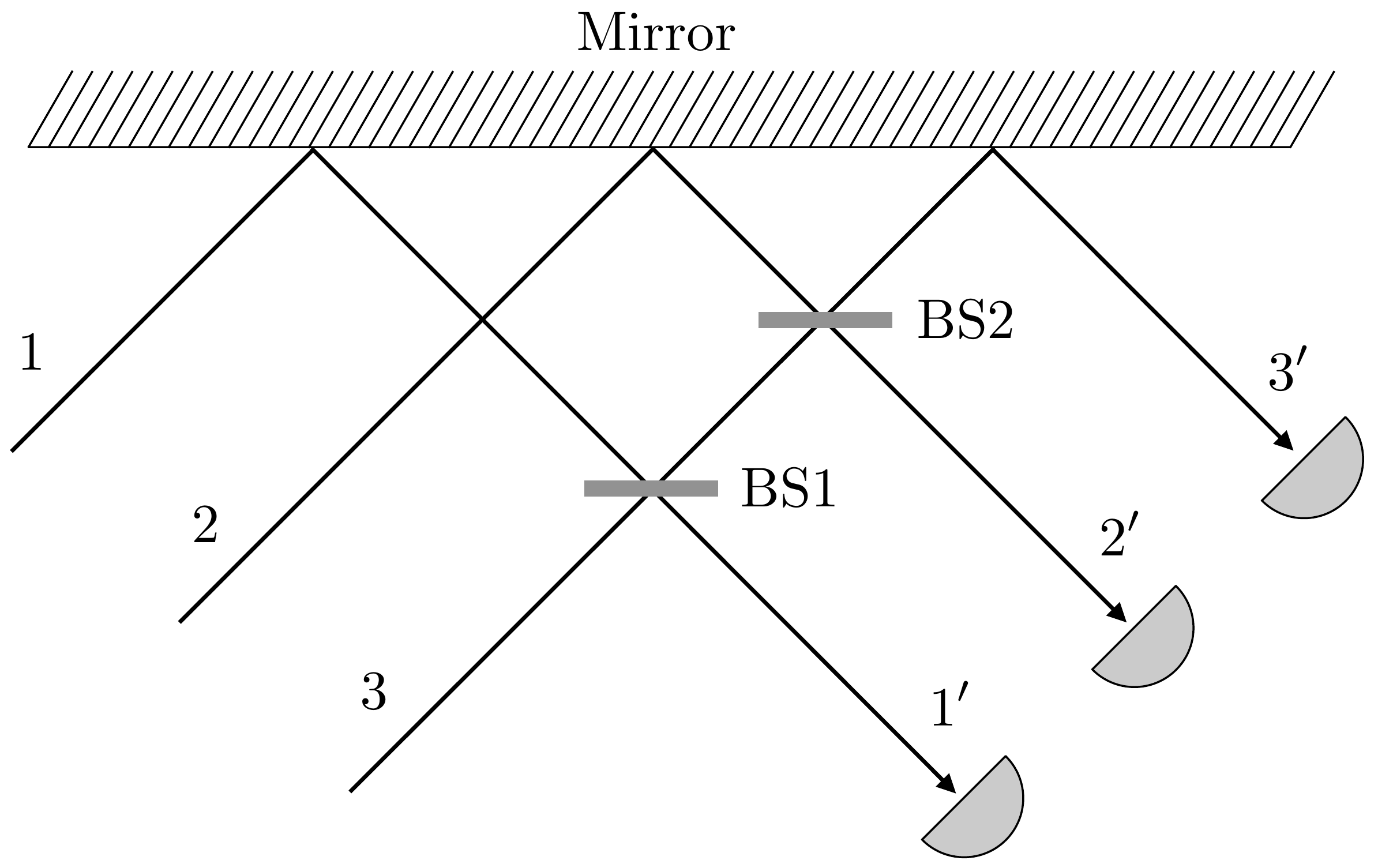}
   \caption{Six-port linear optics implementation of the optimal state separation protocol. The transmission (reflection) coefficients of the beamsplitters, BS1 and BS2 are given by the (off-)diagonal entries of the matrices in Eqs.~(\ref{M1}) and~(\ref{M2}), respectively. The input states are fed through ports $1$ and $2$ as a superposition of zero and one photons in each port. The separated states are output through ports $1'$ and $2'$. Port $3$ at the input is always in the vacuum state. A click in the photodetector placed in port $3'$ signals failure.}
   \label{fig:5}
\end{figure}

The three input ports are labeled \mbox{$1$, $2$, $3$} in the figure. A three-dimensional Hilbert space is spanned by the three orthogonal basis vectors corresponding to one photon in port $i$ and vacuum in the other two ports.  Thus, the basis vectors are $|1\rangle=a_1^\dagger|000\rangle=|001\rangle$,  $|2\rangle=a_2^\dagger|000\rangle=|010\rangle$ and $|3\rangle=a_3^\dagger|000\rangle=|100\rangle$, where~$a_i^\dagger$ is the creation operator of the electromagnetic field in port~$i$, $i=1,2,3$ and $|000\rangle$ is the three-mode vacuum state. Similarly, for the output ports we have~$|1'\rangle=|001\rangle$,  $|2'\rangle=|010\rangle$ and $|3'\rangle=|100\rangle$.

In terms of these basis states, the input states are represented as superpositions of $|1\rangle$ and $|2\rangle$. Note that the third port is always in the vacuum state at the input. Without loss of generality we choose the input states as $|\psi_{1}\rangle=|1\rangle$ and $|\psi_{2}\rangle = s|1\rangle +\sqrt{1-s^{2}}|2\rangle$, and the output states as  $|\psi_{1}^{\prime}\rangle=|1'\rangle$ and $|\psi_{2}^{\prime}\rangle = s'|1'\rangle +\sqrt{1-s'^{2}}|2'\rangle$. Then  Eqs.~(\ref{U1}) and~(\ref{U2}) can be written as
\begin{widetext}
\begin{eqnarray}
U |1\rangle&=&\sqrt{p_1}|1'\rangle+\sqrt{q_1}|3'\rangle,\label{U1 impl}\\[.2em]
U\left( s|1\rangle+\sqrt{1-s^2} |2\rangle\right)&=&\sqrt{p_2}\left( s'|1'\rangle+\sqrt{1-s'^2} |2'\rangle\right)+\sqrt{q_2}|3'\rangle,\label{U2 impl}
\end{eqnarray}
\end{widetext}
which corresponds to the choice $|\phi\rangle|\alpha_0\rangle=|3'\rangle$. The detection of a photon in the output port $3'$ signals that separation failed. The state~$|\psi_2\rangle$ can be produced  in a standard way by sending a photon into a beam splitter with suitable transmission and reflection coefficients.

For simplicity, we consider equal prior probabilities $\eta_1=\eta_2=1/2$, but the same setup can be used in the general case. As mentioned above, for equal priors we must have $q_1=q_2=Q$ and $p_1=p_2=1-Q$ and the unitarity condition Eq.~(\ref{unit cond}) can be solved explicitly. The solution is given by $Q=Q_{-1}$ in Eq.~(\ref{Q's}). Substituting in Eqs.~(\ref{U1 impl}) and~(\ref{U2 impl}) we obtain two columns of the matrix of the unitary transformation $U$ in the basis introduced above. The remaining  column can be easily obtained imposing unitarity. After some algebra we have
\begin{equation}
[U]=\begin{pmatrix}\sqrt{\frac{1-s}{1-s'}} & -\frac{s-s'}{\sqrt{(1-s')(1+s)}} & -\sqrt{\frac{(1+s')(s-s')}{(1-s')(1+s)}}\\[.7em]
0 & \sqrt{\frac{1+s'}{1+s}} & -\sqrt{\frac{s-s'}{1+s}}\\[.7em]
\sqrt{\frac{s-s'}{1-s'}} & \sqrt{\frac{(1-s)(s-s')}{(1+s)(1-s')}} & \sqrt{\frac{(1-s)(1+s')}{(1-s')(1+s)}}
\end{pmatrix}.
\end{equation}
Using~\cite{reck,BergouImp} we can write $U$ as the product $U=M_1 M_2$, where the matrices of $M_1$ and $M_2$  are
\begin{eqnarray}
{}[M_1]&=&\begin{pmatrix}\sqrt{\frac{1-s}{1-s'}} & 0 & -\sqrt{\frac{s-s'}{1-s'}}\\
0 & 1 & 0\\
\sqrt{\frac{s-s'}{1-s'}} & 0 & \sqrt{\frac{1-s}{1-s'}}
\end{pmatrix},\label{M1}\\[1em]
{}[M_2]&=&
\begin{pmatrix}1 & 0 & 0\\
0 & \sqrt{\frac{1+s'}{1+s}} & -\sqrt{\frac{s-s'}{1+s}}\\
0 & \sqrt{\frac{s-s'}{1+s}} & \sqrt{\frac{1+s'}{1+s}}
\end{pmatrix}.
\label{M2}
\end{eqnarray}
We immediately recognize that the transformation $M_1$ and $M_2$ can be implemented with beamsplitters, labeled in Fig.~\ref{fig:5} by  BS1 and BS2, respectively. The corresponding matrix elements provide the transmission (diagonal) and reflection  (off-diagonal) coefficients of these beamsplitters.

The degree of separation attained by the protocol can be certified by statistical analysis of the photon counts in the detectors placed in the ports $1'$ and $2'$, whereas those in the detector placed in port $3'$ provide the failure rate~$Q$. 

Alternatively, one might consider the transformation provided by the set-up as a subroutine, probabilistically performing the requested state separation, as part of a larger protocol. One can achieve this by removing the detectors in $1'$ and $2'$ and feeding the output states into some subsequent unit for further processing. Hence, this implementation can be thought of as a separation module in a larger set-up.

\vspace{2em}

\section{Conclusions and Outlook}\label{conclusion}

In this paper we have addressed quantum state separation for two known pure states with arbitrary prior probabilities. The degree of separation required by a probabilistic transformation determines its minimum failure rate. Thus, knowing the relationship between these quantities  for arbitrary priors and arbitrary overlap of the input states is a valuable piece of knowledge  for quantum information processing. It provides the ultimate limits on the processing of information allowed by nature and sets the performance scale for experimental implementations of such processing protocols. 

We have given a full account of state separation by focusing separately  on the various situations that one may encounter in quantum state processing. We first dealt with the optimization of protocols that have a fixed degree of separation, such as probabilistic perfect cloning. We  have revisited, completed and extended our results in~\cite{us1}. We have also given some technical details that were missing there. We have next considered the optimization of protocols for which a maximum allowed failure rate, or margin, is given. We have computed the maximum separation that a state transformation can possibly achieve as a function of the overlap of the input states and we have characterized the tradeoff between separation and failure rate for fixed initial overlap.

We have shown that a phenomenon analogous to a second order symmetry breaking phase transition arises in the limit of full separation, when the processed states become orthogonal. We have characterized it in the various situations discussed in the previous paragraph. Similar phase transitions have been discussed in connection with unambiguous discrimination of two or more states. The phenomenon arises from the high non-linearity of the unitarity constraints imposed by quantum mechanics.  

We have approached the optimization problems discussed in this paper from a geometrical viewpoint that enabled us to gain a great deal of intuition about the solutions. This intuition has been the guiding line towards finding analytical results. Although a closed form for the solutions does not exist in the general case because of the high-degree non-linearity of the problem, our approach provides all the required relations between the relevant quantities in a clear and detailed way. The same geometrical approach has been applied in~\cite{us1}  and~\cite{Bergou1} where it proved equally powerful, and it can be applied to other optimization problems in quantum information processing where similar highly non-linear constraints arise. In this direction, we have some work in progress on probabilistic approximate cloning of two states and perfect cloning of three states. 

\begin{acknowledgments}
This publication was made possible through the support of a Grant from the John Templeton Foundation. The opinions expressed in this publication are those of the authors and do not necessarily reflect the views of the John Templeton Foundation. Partial financial support by a Grant from PSC-CUNY is also gratefully acknowledged. The research of EB was additionally supported by 
the Spanish MICINN, through contract FIS2013-40627-P, the Generalitat de
Catalunya CIRIT, contract  2014SGR-966, and ERDF: European Regional Development Fund. EB also thanks the hospitality of Hunter College during his research stay.
\end{acknowledgments}

\end{document}